\documentstyle[12pt]{article}
\begin{document}

\def\beqra{\begin{eqnarray}} \def\eeqra{\end{eqnarray}}
\def\beqast{\begin{eqnarray*}} \def\eeqast{\end{eqnarray*}}
\def\beq{\begin{equation}}	\def\eeq{\end{equation}}
\def\be{\begin{enumerate}}   \def\ee{\end{enumerate}}
\def\fnote#1#2{\begingroup\def\thefootnote{#1}\footnote{#2}\addtocounter
{footnote}{-1}\endgroup}

\vspace{24pt}

\begin{center}
{\bf Structure and Parametrization of Stochastic Maps of Density Matrices}

\vspace{48pt}

E.C.G. Sudarshan\\
Department of Physics and Center for Particle Physics\\
University of Texas, Austin, Texas 78712-1081\fnote{*}{e-mail: 
sudarshan@physics.utexas.edu}
\end{center}

\vspace{36pt}

\abstract{ The generic linear evolution of the density matrix of a system with a 
finite-dimensional state space is by stochastic maps which take a density matrix
linearly into the set of density matrices.  These dynamical stochastic maps form a
linear convex set that may be viewed as  supermatrices.  The property of
hermiticity of density matrices renders an associated supermatrix hermitian and
hence diagonalizable; but the positivity of the density matrix does not make this
associated supermatrix positive.  If it is positive, the map is called completely
positive and they have a simple parametrization.  This is extended to all positive
(not completely positive) maps.  A contraction of a norm-preserving map of the
combined system can be contracted to obtain all dynamical maps.  The
reconstruction of the extended dynamics is given.}

\newpage

\baselineskip=24pt

\parindent2.5em

\section{Introduction:  Dynamical Maps}

The Quantum Density Matrix $\rho$ is the statistical state and therefore quantum 
dynamics is the evolution of the density matrix.  For a closed system this
evolution is by a unitary time-dependent operator:
\beq
\rho(t) = U(t)\,\rho(0)\,U^{\dag}(t).
\eeq
The evolution is linear.  But if we have an open system, its dynamics cannot be by
a unitary evolution but a more general linear evolution[1]:

$$\rho(t)=A(t)\,\rho(0)$$
where $A$ is a linear map of the density matrix into a density matrix.  This 
superoperator $A(t)$ can be written as a supermatrix.
\beq
\rho_{r,s}(0)\longrightarrow A_{rs,r's'}(t)\rho_{r's'}(0)=(A(t)\rho )_{rs}.
\eeq

Then the supermatrix $A(t)$ must satisfy the following three constraints:
\beqra
A_{sr,s'r'}(t) &=& A_{rs,r's'}(t) \nonumber \\
\rho(0)\geq 0 &\longrightarrow & A_{rs,r's'}\rho_{r's'}\geq 0 \\
A_{sss,r's'} &=& \delta_{r's'}. \nonumber
\eeqra
These are consequences of the hermiticity, positivity and trace properties of the 
density matrices.  These properties can be best expressed in terms of the
dynamical matrix[2].
\beqra
B_{rr',s's} (t) &=& A_{rs,r's'}(t) \nonumber \\
B_{s's,rr'}^*(t) &=& B_{rr,s's}(t) \nonumber \\
\!\!\!x^*_ry_r, && \!\!\!\!\!\!\!\!\!\!\!\!\!\! B_{rr',s's}(t) x_s\,y_{s'}^* \geq
0    \nonumber
\\ B_{nr',s'n}(t) &=& \delta_{r's'} \;.
\eeqra

Thus $B$ is a hermitian matrix which gives nonnegative expectation value for 
supervector that can be factorized:
\beq
u^{\dag}\, B\,u \geq 0~~\mbox{if} ~~ u_{rs} = x_r y_s^*\,.
\eeq
It is sufficient if $B$ is nonnegative but it is not necessary.  If $B\geq 0$,
then we  will call the map ``completely positive;"[2]  but if only the positivity
condition (4) is satisfied, we will call the map ``positive but not completely
positive".  Since $B$ is hermitian, according to (4) it follows that it has an
eigenvector decomposition
\beq B_{rr',s's} = \sum \mu_\alpha\, \zeta^{(\alpha)}_{rr'}\,\zeta^{*(\alpha)} =
\zeta\,M\zeta^{\dag}
\eeq
when $M$ is the diagonal matrix with eigenvalue $\mu_{\alpha}$, and
$\zeta_{rr'}^{(\alpha)}$ are the normalized  eigenvectors.  For a completely
positive map all the $\mu_{\alpha}$ are nonnegative, but this will not be true for
not completely positive maps.  If all the $\mu_{\alpha}$ are nonnegative, we can
absorb them by defining the eigenvector
$$ 
C^{(\alpha)}_{rr'} = \mu_\alpha^{ 1/2} \;\zeta_{rr'}^{(\alpha)}.
$$
So for, a completely positive map[2]
\beq
\rho\longrightarrow \sum_\alpha \, C^{(\alpha)}\rho C^{(\alpha)\dag} 
\eeq
with the trace condition
\beq
\sum_\alpha\, C^{(\alpha)\dag} C^{(\alpha)}=1\,.
\eeq

\section{Not Completely Positive Maps}

For a not completely positive map, some eigenvalues $\nu$ are negative.  We may
then  define
\beq
\rho \longrightarrow \sum_\alpha\, C^{(\alpha)\dag}\, \rho\; C^{(\alpha)}
-\sum_\beta
\,D^{(\beta)\dag}\rho\;D^{(\beta)}
\eeq
with
\beq
D_{rr'}^{(\beta)} =(|\nu|)^{1/2}\eta^\beta_{rr'}\;.
\eeq
It will turn out that the number of negative eigenvalues cannot exceed the number 
of positive eigenvalue.  The trace condition now becomes
\beqra
\sum_\alpha\, C^{(\alpha)\dag}\, C^{(\alpha)} &-& \sum_\beta\,
D^{(\beta)\dag}\,D^{(\beta)} =1 \;. \\[6pt]
\sum_\alpha C^\alpha u_\alpha u_\alpha^{\dag} \,C^{\alpha\dag} &-& D^{\beta)}
u_\beta \,u_\beta^{\dag}\;D^{(\beta)\dag} \geq 0\;.
\eeqra
In dealing with maps we have the convexity property:
\beq
B^I\;\cos^2\theta + B^{II}\, \sin^2\theta = B
\eeq
is a positive map if $B^I$ and $B^{II}$ are positive maps.  Every map that cannot
be  expressed in this manner is called extremal:
$$ B=B^I\,\cos^2\theta + B^{II}\, \sin^2\theta\longrightarrow B^I\equiv B^{II}\,. 
$$
If we can determine all extremal maps, we can generate all maps from them.

Since the density matrices are restricted by positivity and the trace condition, 
they form a compact set.  The dynamical maps map from a compact set into a compact
set, the maps also form a compact set.[1]  Such a set can be generated from its
extremal elements.  Then
\beq
\sum_\alpha \,{\rm{tr}}\, (C^{(\alpha)}\, C^{\dag}) - \sum_\beta{\rm{tr}}\,
(d^{\beta)} d^{(\beta)\dag})=N\;. 
\eeq
Define
\beq
\sum_\alpha\, C^{(\alpha)}\,C^{(\alpha)\dag} = J\geq 0\; ,\; \sum_\beta\,
D^{(\beta)\dag} = K\geq 0\,.
\eeq
We can do unitary transformations $U,V$ on the initial and final density matrices 
we can diagonalize  $J$.  By virtue of (11) this will diagonalize $K$.  Then

$$J-K=1$$

If the eigenvalues of $K$ are $k^2_\alpha$ and for $J$ they are $j^2_\alpha$, then
$$
 k^2_\alpha = j^2_\alpha +1\,.
$$
Define $\vartheta_\alpha$ so that
\beq
j_\alpha=\sinh \vartheta_\alpha ~,~ k_\alpha=\cosh\;\vartheta_\alpha\;.
\eeq
Define
\beqra
C^{(\alpha)} &=& \cosh \vartheta_\alpha\;M (\alpha) \nonumber \\
D^{(\alpha)} &=& \sinh \vartheta_\alpha\;N (\alpha) 
\eeqra
so that
\beq
\sum^m_{\alpha=1}\,M_{(\alpha)}\; M^{\dag}_{(\alpha)}=1~,~\sum^n_{\beta=1}\,N
(\beta) N^{\dag} (\beta)=1\;.
\eeq
Parametrizising these matrices is already known[3].  In those cases where $\sinh
\vartheta_\alpha=0$, the matrix 
$N(\alpha)$ are not defined and we sum over a smaller set than the set of
$M(\alpha)$.

Choose a unitary transformation $W_1$ so that $M^1_{(1)}$ is diagonal with
eigenvalues
$\cos\,\theta^{(1)}_1$:
\beq
W^{\dag}_1\;M(1) \,W_1=\left( \begin{array}{ccc}            
\cos\,\theta^{(1)}_1 &&0 \\
&\cos\,\theta^{(1)}_2 \\
0&& \ddots \cos^2\,\theta_m^{(1)}\end{array}\right)
\eeq
Than
\beq
W^{\dag}_1\; \sum^m_2\; C^{(\alpha)}\, C^{(\alpha)\dag} \, W_1= \left(
\begin{array}{ccc}
\sin^@\,\theta^{(1)}_1 &&0 \\
&\sin^2\,\theta^{(1)}_2 \\
0&&\ddots \sin^2\,\theta_n^{(1)}\end{array}\right)
\eeq
Define
\beq
M^{(2)}_{(\alpha)} = \left( 
\begin{array}{ccc}
\sin\,\theta^{(1)}_1 &&0 \\
&\sin\,\theta^{(1)}_2 \\
&&\ddots \sin\,\theta_N^{(1)}\end{array}\right) M(\alpha),\;2\leq\alpha\leq N\,.
\eeq
Then define the $(N-1)\times(N-1)$ matrices 
which satisfies
$$ M^{(2)}(m)\;M^{(2)\dag}(m)=1_{N-1\times N-1}$$ 
We now repeat the procedure.  Define
$$
W^{\dag}_2\,M^{(2)}(2)\,W_2= \left( \begin{array}{ccc}
\cos\,\theta^{(2)}_1 &&0 \\
&\cos\,\theta^{(2)}_2 \\
&0 & \ddots \cos\,\theta_m^{(2)}\end{array}\right)
$$
and a new orthogonal transformation $\sigma^{(2)}_{\alpha}$ such that
$\theta^{(2)}_2=0$ and we can define
$M^{(2)}_{(m)}$ so that  we can introduce the $m-2$ matrices $M^{(2)}(m)$ which
satisfy 

\beqra
M^{(2)} (m) &=& \left( 
\begin{array}{ccc} \cos\,\theta_1^{(3)} &&0 \\
&\cos\,\theta_2^{(3)} &\\
0 & & \ddots \cos^{(3)}\,\theta_3\end{array} \right) \\  
\sum\, M^{(3)}_(m) M^{(3)\dag}_{(m)} &=& 1_{N,\,N} \nonumber
\eeqra

This procedure can be carried out until all matrices are parametrized.  The $M_x$ 
needs $m$ parameter $\theta^{(1)}_1\ldots, \,\theta^{(1)}_m $ and so on until we
get, a total of
$ m^2$ angles.  The unitary matrices are also relevant parameters which
determine the dynamical map.

The same procedure can be carried out to parametrize $K$.  We need $n^2$ angles.  
Together then, the matrices $J,K$ together require angles
$\vartheta_1\,\ldots,\,\vartheta_m\; , \; m^2$ angles for
$J$, $n^2$ angles $K$; a total of $(m^2+n^2)$ parameters to determine the matrices
$C^{(\alpha)},~D^{(\beta)}$ to within
$m+n$ unitary matrices according to
$$
C^{(\alpha)}\longrightarrow C^{(\alpha)}\;U^{(\alpha)}
$$
which leave (11) unchanged but change the maps $B_{rr',s's}$.

\section{Dynamical Maps as Contractions:}

A straightforward way[4] of generating positive maps is to consider a unitary 
evolution of a coupled system.  $R,S$ with $S$ being the system and $R$ a
`reservoir'.  If we take a direct product density matrix
$$
\rho\rightarrow \rho\times\tau\rightarrow V\rho\times\tau\,V^{\dag}
$$
where $V$ is a unitary matrix in the direct product space
$$H_m \times H_n.$$
Then
\beq 
\rho_{rs}\times \tau_{ab}\rightarrow V_{ra,r'a'}\,
\rho_{r's'}\tau_{a'b'}\;V_{sb,s'b}^*\,.  
\eeq
The partial trace operation is a contraction
\beq
V\;\rho\times\tau\,V^{\dag} \rightarrow V_{rn,r'a'}\,\rho_{r's'}\tau_{a'b'}\;
V_{sb,s'b'}
\eeq
If $\tau$ is made diagonal (if necessary by a unitary transformation in $H_n$) and 
the eigenvalues are $\tau_{(1)},\, \tau_{(2)}\, ,\ldots, \,\tau_{(n)}$, the map is
\beq
\rho_{rs}\longrightarrow \sum_{{\nu \atop  r's'}\atop n} V_{rr'} (n,\nu,r)
\,\rho_{r's'}\,\tau(\nu)\,V_{ss'}^*\, (n,\nu,\tau).
\eeq
Clearly, if $\tau$ has more than one nonzero eigenvalue, the map is not extremal.  
So as far as extremal maps are concerned
\beq
\rho_{rs}\longrightarrow\sum_n\, V_{rr'}(n)~\rho_{r's'}\,V^*_{ss'}(n)
\eeq
which is of the standard form [3]
$$
 \rho \longrightarrow \sum_\alpha \, C^{(\alpha)}\,\rho\,C^{(\alpha)\dag}
$$
in which $n^\alpha$ runs over at most $1\leq n\leq m $.  We note that all these
maps are  completely positive.  We can also do an inverse reconstruction:  given an
extremal completely positive map
$$
\rho\longrightarrow \sum\, C^{(\alpha)}\,\rho\,C^{(\alpha)\dag}
$$
we can define a reservoir matrix $\tau$ with all elements except $\tau_{11}$  
zero and $\tau_{11}=1$.  Then we can construct a unitary matrix $V$ in $mn$
dimensions with:
\beq
V_{r\alpha,r'1} = C_{rr'}(\alpha)\;.
\eeq
The conditions on $C(\alpha)$ are transcribed into
\beq
\sum_\alpha\;V_{r\alpha,r'},\; V^*_{s\alpha,s'},\, = \delta_{rs}
\eeq
which is necessary for $V$ to be a unitary matrix.  The ambiguity in constructing 
other elements does not affect the map.  Thus a completely positive extremal map
can always be obtained as a contraction of a unitary evolution[6].  The other
elements $ C^{(\alpha)} $ can be used to carry out the construction of the matrix
$V$ and a generic diagonal matrix $\tau$.

\section{Not Completely Positive Maps as Contractions}

What about not completely positive maps?  To obtain such a map by contraction we 
generalize the auxiliary space $H_N$ to be a space with an indefinite metric and
$V$ to be a pseudounitary operator in the $MN$ dimensional space.  Positivity is
guaranteed if the generalized density matrix is entirely within the convex of
positive metric states of the $MN$ dimensional space.  Then the sum over $n$ in
(26) goes over both positive and negative metric terms; but the resultant density
matrix is nonnegative.

We can invert this derivation to realize the most general extremal not completely 
positive map as the contraction of a larger evolution in an indefinite metric
space for the reservoir and a density matrix $\tau$ of the reservoir system to
have a single eigenvector (with positive metric) with eigenvalue unity, the others
being trivial.

Since such reservoirs are somewhat artificial, we may consider this reconstruction 
as a purely formal device.
\newpage

\vspace{18pt}
\noindent
\Large
{\bf Summary}

\vspace{18pt}
\normalsize

We have studied linear dynamical maps which take the set of density matrices into 
the set of density matrices.  These maps form a convex set and are compact in the
case of completely positive maps.  The search for extremal maps gives us the
restriction that we need at most $N$ terms for an extremal map.  Those maps can
be obtained as contractions of a direct product system:  the extremal maps
correspond to a reservoir matrix which is a projection.  Conversely we can
reconstruct the unitary evolution of the expanded system from the map itself.

The considerations are extended to positive but not completely positive dynamical 
maps.  The  extremal maps still contain at most $N$ terms.  We can obtain these as
contractions of an extended system with a pseudounitary evolution matrix.  We
could also reconstruct the extended pseudounitary evolution from the maps.

In the systematic parametrization we find we need $N(N-1)/2$ parameters for the 
completely  positive map apart from a set of unitary $N \times N$ matrices.  For
the corresponding parametrization of the not completely positive maps we have
$(m^2 + n^2)$ for $m$ positive and
$n$ negative eigenvalues for the dynamical matrix and the unitary matrices (or
less).

These results generalize the results obtained two decades ago by Gorini and 
Sudarshan[6] for $2 
\times 2$ matrices.

Needless to say, however complicated the dynamical processes leading to the linear 
stochastic  evolution that is represented by the dynamical map, we see that the
same dynamics obtains with a reservoir having dimension $N^2 \times N^2$.  In the
case of an extremal map it suffices to have a reservoir with a state space of the
same $N \times N$ dimensional density matrices.

In this paper we have only dealt with dynamical maps, not the continuous semigroup 
of  evolutions.  This study was carried out by A. Kossakowski[9] and followed by
others.[8,9]  In these while the semigroup generators are parametrized no attempt
is made to embed them in a larger system.  Since the Zeno effect[10] operates for
very small time intervals, care must be taken in generating a semigroup from the
dynamics of an extended system.  We hope to examine this question in the near
future.

\end{document}